# Interpretable inverse design of particle spectral emissivity using machine learning

Mahmoud Elzouka[1], Charles Yang[1,2], Adrian Albert, Sean Lubner[1*], Ravi S. Prasher[1,2*]

[1]Energy Storage & Distributed Resources Division, Lawrence Berkeley National Laboratory, Berkeley, CA 94720, USA

[2]Department of Mechanical Engineering, University of California, Berkeley, CA 94720, USA

* Corresponding authors: slubner@lbl.gov, rsprasher@lbl.gov

**Abstract:**
We examine the optical properties of a system of nano and micro particles of varying size, shape, and material (including metals and dielectrics, and sub-wavelength and super-wavelength regimes). Training data is generated by numerically solving Maxwell's Equations. We then use a combination of decision tree and random forest models to solve both the forward problem (particle design in, optical properties out) and inverse problem (desired optical properties in, range of particle designs out). We show that on even comparatively sparse datasets these machine learning models solve both the forward and inverse problems with excellent accuracy and 4 to 8 orders of magnitude faster than traditional methods. A single trained model is capable of handling the full diversity of our dataset, producing a variety of different candidate particle designs to solve an inverse problem. The interpretability of our models confirms that dielectric particles emit and absorb electromagnetic radiation volumetrically, while metallic particles' interaction with light is dominated by surface modes. This work demonstrates the possibility for approachable and interpretable machine learning models to be used for rapid forward and inverse design of devices that span a broad and diverse parameter space.

**Main Text:**
Controlling light-matter interactions is crucial for a variety of important applications including energy harvesting[1,2], radiative cooling[3,4], heating[5], and computing[6–8]. Engineering these optical properties for most practical applications requires inventing new materials or designing novel geometries with nano- or micro-scale features. Given a known geometric design of known materials, its optical properties can be deterministically calculated by numerically solving Maxwell's Equations for virtually any problem[9]. This amounts to solving the so-called "forward problem." However, in most practical applications a desired set of optical properties is known while a geometric design must be found to produce those properties. It is very difficult to solve this "inverse design problem" and find appropriate geometric designs from a given set of desired optical properties[10]. In general, the inverse design problem is nonlinear and has a one-to-many mapping. As a consequence, the standard solution approach is to iteratively solve the forward problem using trial-and-error with various optimization techniques[11]. Good initial guesses and computation resources are necessary but not always sufficient for the optimization to be tractable.

Machine learning has been proposed to create surrogate models that can return approximate solutions for the forward problems fast, which consequently accelerates the iterative design optimization process. More interestingly, machine learning has also shown some success in further accelerating the inverse design process by performing the inverse mapping in a single-shot (i.e. without iteratively solving the forward problem) for some optics problems (e.g., concentric spherical shells[12] and periodic metasurfaces[13–16]). However, these approaches succeed in part by restricting the scope of considered geometries and materials to be very narrow and homogeneous, limiting their wider utility. Furthermore, these approaches typically lack interpretability and explainability of the ML models. So while an algorithm may come up with a plausible design, the reasoning underlying the design is hidden from the researcher. This is primarily due to the frequent use of Deep Neural Networks (DNN). Interpreting DNN's is difficult; it is still an open research problem[17]. Consequently, it is difficult for an engineer to learn design principles from most inverse designs generated by a DNN.

In this work we solve both the forward and inverse optical design problems for solid particles spanning a variety of different geometry classes, materials, and size regimes, using a single and intrinsically interpretable ML model (decision trees). Our dataset consists of spectral emissivity curves numerically calculated for 15,900 particles of varying shape (spheres, parallelepipeds, triangular prisms, and cylinders), aspect ratio, size (nanometers to tens of microns; spanning sub-wavelength to super-wavelength regimes), and material ($SiO_2$, SiN, and Au). Given the dimensionality of this parameter space this number of data points corresponds to quite sparse coverage; for a given material + shape combination each linear feature dimension (e.g. particle length) typically varies by more than 2 orders of magnitude over its range but is sampled on average by only 9 data points (given our test/train split of 50/50). The emissivity is calculated by direct numerical simulation (DNS) of Maxwell's Equations (see Methods section).

We display our entire dataset in **Figure 1** (low opacity points). The spectrally integrated emissivity of each particle is plotted against its area-to-volume ratio, which represents the particle size. We note that the dielectrics show clear correlations between emissivity and particle size. This is explained by the volumetric nature of emissivity for dielectric materials, which exhibit low attenuation[20]. On the other hand, the emissivity of metals has no clear correlation with particle size, other than it is more diverse for smaller particles. This is explained by the dominating effects of surface and localized electromagnetic modes that can be supported by small metal particles (e.g., localized plasmons), which can significantly influence the emissivity. These modes depend on surface geometry more so than the overall particle size. By including materials with different emission mechanisms we aim to demonstrate the flexibility of our machine learning model to handle diverse datasets. Further information on the dataset descriptors, dataset generation, and model implementation is given in the Methods section and in **Fig. S2-5**.

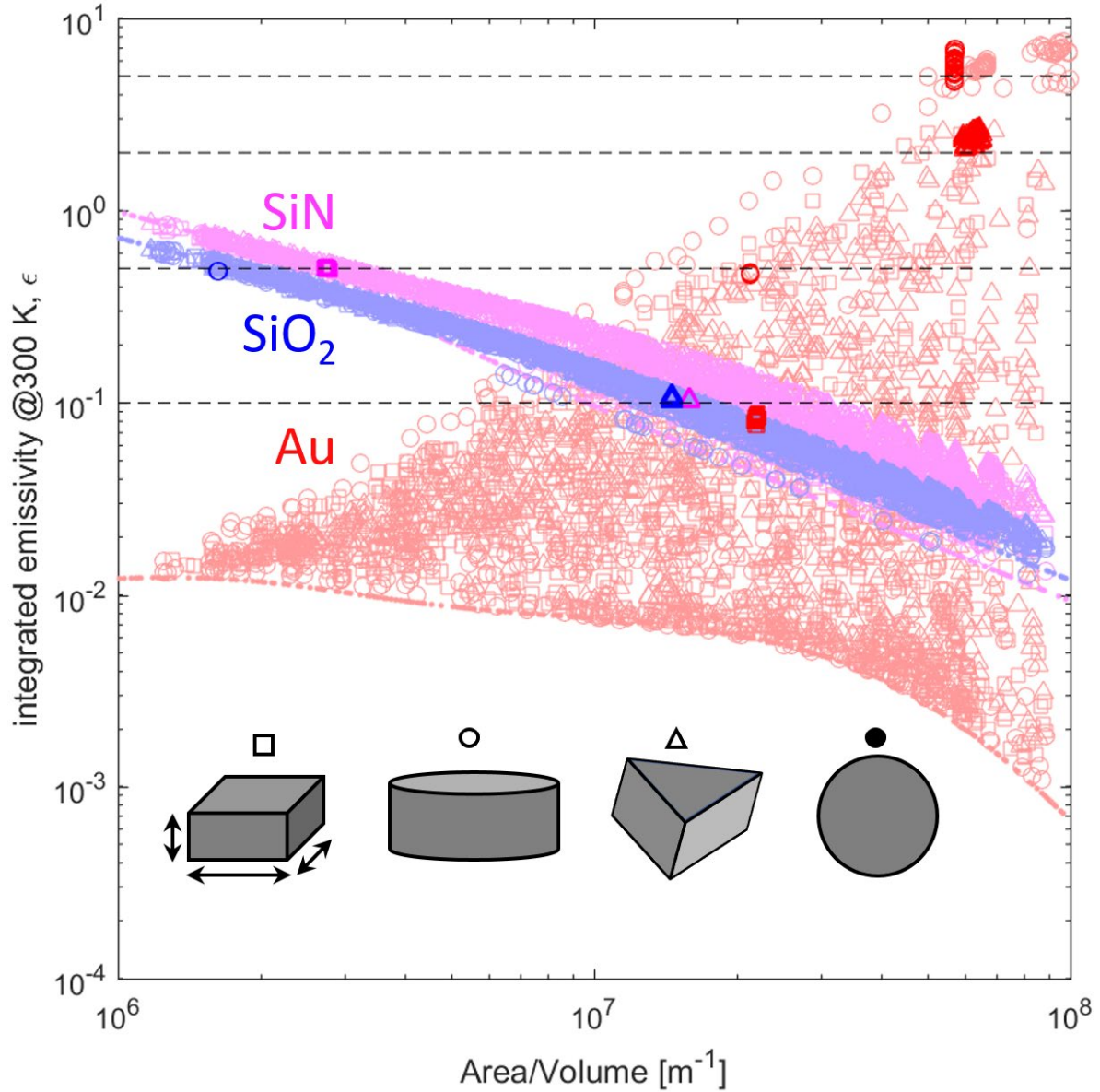

**Figure 1**. Spectrally and hemispherically integrated emissivity at 300 K for every particle in our training dataset (low opacity points) as a function of particle area-to-volume ratio. Colors and symbol shapes represent different materials and geometries, respectively. Dashed lines and fully opaque points are target emissivities and their corresponding model-generated designs, respectively (discussed later in the text).

We first solve the forward problem of predicting particle emissivity by using a Random Forest (RF)[18], which is an ensemble of decision trees (DT). These models average and mix data during the training process so that they can predict emissivity spectra that they have never seen during training. For training, each particle is represented to the model as a length-6 array of parameterized geometric features and a material type. In particular, these 6 features are: the mutually orthogonal shortest, middle, and longest dimensions of the particle, the area-to-volume ratio, and one-hot encodings of the geometry class and of the material type. We separately train two RF models using the same data: one to predict the total spectrally integrated emissivity (scalar target) and the other to predict the full spectral emissivity spectrum (vector target). Each training data point is therefore a length-6 input array of descriptor features with a corresponding output scalar value (spectrally integrated emissivity) or output vector (emissivity spectrum spanning near to far infrared). We use a 50/50 test/train split (i.e. train using 7,950 particles and test on the remaining different 7,950 particles). The model's performance on the test set is shown in **Figure 2** for both scalar and vector targets. The model errors (**Figure 2b** insets) vary across different materials and geometries, but are always small (generally <10%). **Figures S6 and S7** show more detail of the RF inference error broken down by geometry and material, respectively. Higher error for Au compared to $SiO_2$ and SiN is consistent with the greater diversity of optical interactions that can occur on metallic particles due to resonant modes, as can be seen in **Figure 1**.

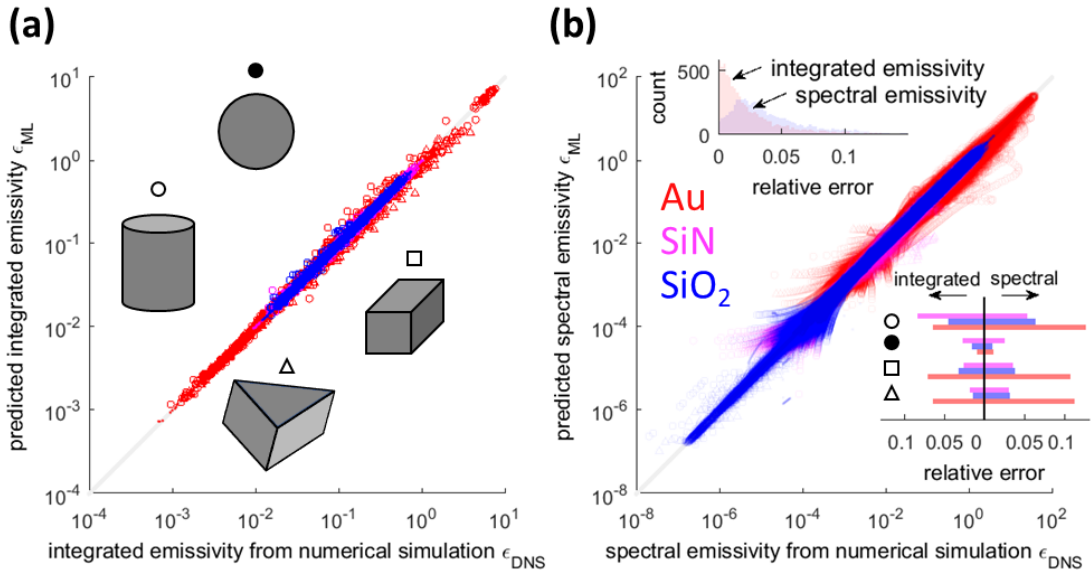

**Figure 2**. Machine learning predicted ($\epsilon_{ML}$) compared to direct numerical simulation results ($\epsilon_{DNS}$) for integrated emissivity (a) and spectral emissivity (b). Colors and symbol shapes represent different materials and geometries, respectively. Upper inset represents the distribution of relative error (defined in the Methods section) of the random forest model. The lower inset shows the relative error by material and geometry, averaged over 100 different trained models with different random test/train splits.

Full emissivity spectra predictions for individual particles representing different particle materials and geometry classes are shown in **Figure 3**. Note that these spectra were chosen from among the worst 20% of predictions (as measured by relative error) from the test set for each material and geometry class combination. RF performs well across all geometries and materials.

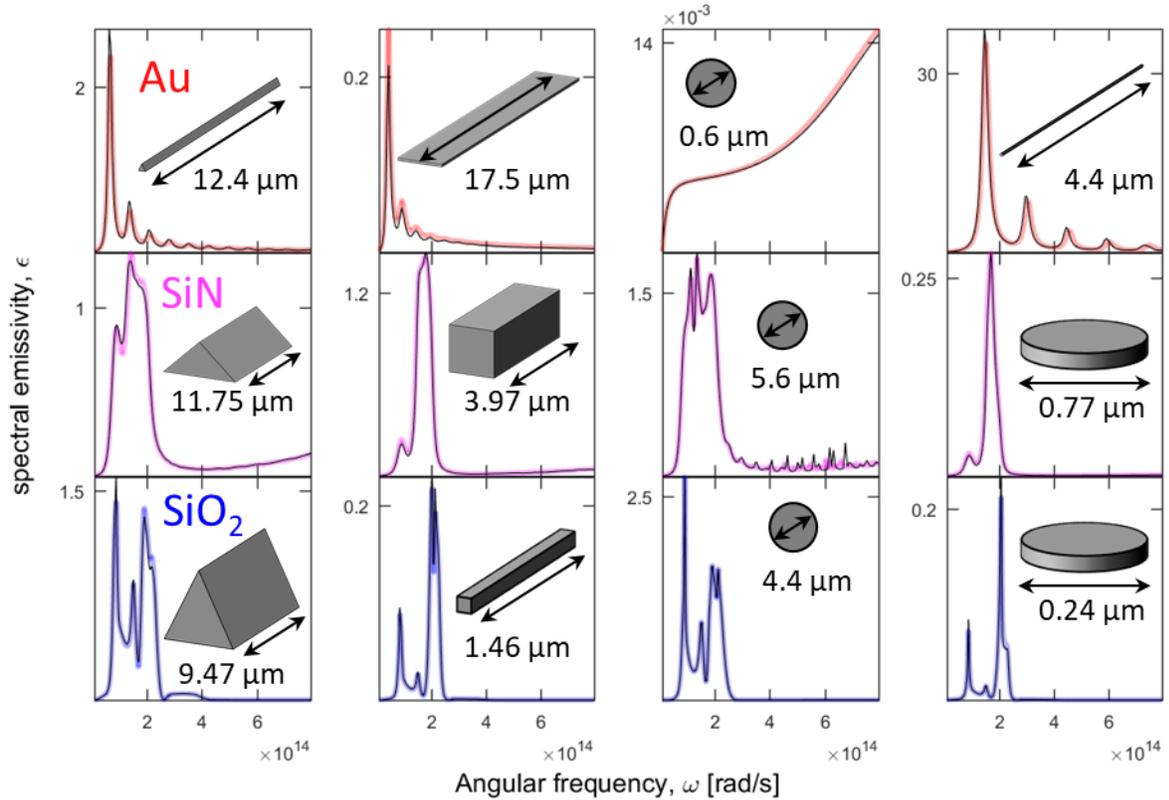

**Figure 3**. Model predictions (colored lines) compared to numerical solutions (black lines) of spectral emissivity for particles of different shapes (columns) and materials (rows). These examples were chosen from among the worst 20% of predictions from the test set.

Next, we wish to solve the inverse design problem. In general random forests offer superior model accuracy and robustness as compared to a single decision tree[18]. However, while decision trees are individually interpretable, they are more difficult to interpret when bootstrapped into an ensemble such as a Random Forest. Additionally, while it is possible to retrace back up a decision tree to perform inverse design, this is not possible for a random forest because its output is an averaged ensemble vote. We correct this reduced interpretability and solve the inverse problem by employing a Combined Multiple Models (CMM) method[19]. A CMM method compresses ensemble-based models (e.g. RF) into a single base model (e.g. DT) without significantly affecting model performance, hence restoring the ability to interpret the model and use it for inverse design. The use of the CMM algorithm has been largely limited because it requires problems for which it is cheap to generate large amounts of synthetic unlabeled training data. For instance, it is difficult to programmatically generate images of faces or dogs or cats, which are canonical computer vision datasets. However, in our inverse design problem we can easily generate a diverse set of particle samples simply by randomly sampling over the ranges of (self-consistent) physical and geometric parameters.

To apply the CMM method we first generate an additional 250×(dataset size) number of random, unlabeled particle designs and use our trained RF to label them (i.e. predict their emissivities). We then train a new decision tree on this larger generated dataset. We call this new decision tree DTGEN because it is trained on generated data. DTGEN is in general a much larger decision tree than DT because it is trained on a much larger dataset. Note that training DTGEN does not require any new real data (i.e. data labeled by DNS of Maxwell's Equations). DTGEN emulates the performance of RF while preserving the interpretability and retracability of a regular decision tree.

To perform inverse design we find the output label ("leaf") on DTGEN whose value (can be a scalar or vector) corresponds closest to the desired optical properties we want our particle to produce. We trace up the decision tree branch from this leaf taking the intersection of all branch-splitting criteria on all design features encountered along the way. The result is a set of design rules for each feature that produces the desired target optical properties. For example, a set of design rules might stipulate a range of lengths for the particle's middle dimension or a subset of acceptable materials or geometry classes. Having a set of design rules naturally captures the "one-to-many" mapping behavior of inverse design problems. We randomly sample self-consistent particle designs from these design rules and calculate their true optical properties using the same DNS scheme for Maxwell's Equations that we used to generate the original training data. The optical properties of these generated designs are then compared to the original target optical properties.

**Figure 4** shows the distribution of spectrally integrated emissivity values produced by the designs generated by DTGEN and DT for a few select target values (target values are labeled on the *x*-axis). To demonstrate the robustness and flexibility of our model, we require that the algorithm generate particle designs using each of the three materials for those target emissivities that are < 1 (because generally only metal structures have apparent emittance > 1). These same DTGEN-produced designs are also highlighted in **Figure 1** (full opacity points are generated designs; dashed black lines are the target values). Both DT and DTGEN perform well at solving the inverse design problem for integrated emissivity, with DTGEN slightly outperforming DT. The corresponding design rules for each sample of DT and DTGEN are shown in **Fig. S10.** In most cases DTGEN suggests stricter design rules than DT, which explains the difference in performance and the tighter distribution of DTGEN-generated designs in **Figure 4**. Note that the model can still accurately solve the inverse design problem in regimes where there is comparatively very little training data (e.g. at high integrated apparent emittance values).

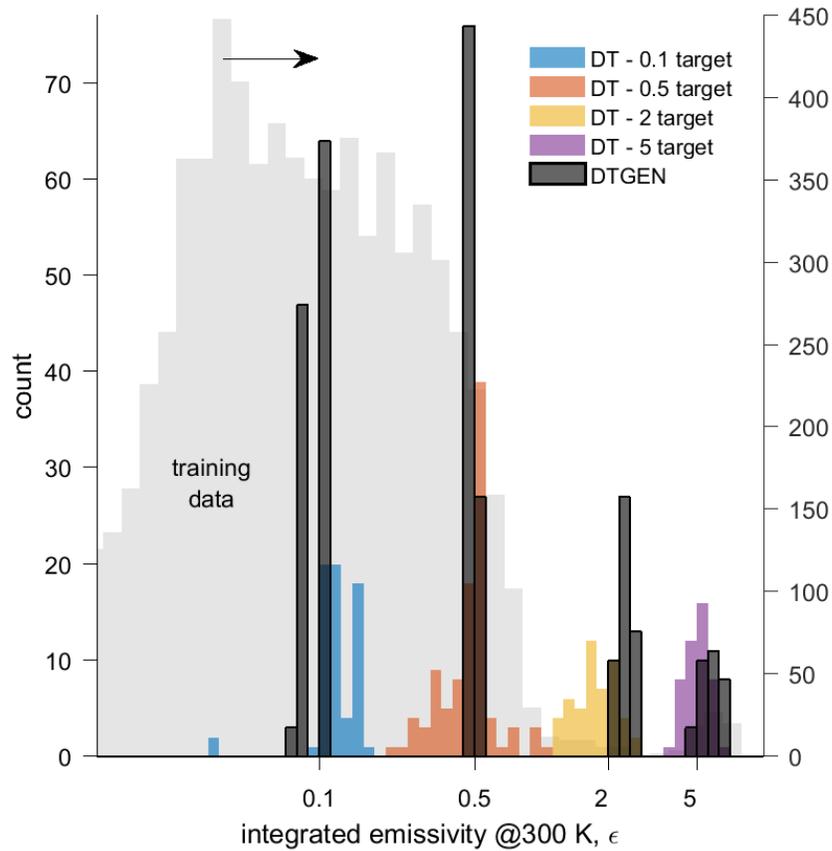

**Figure 4**. Inverse design solutions for integrated emissivity. Target emissivities are 0.1, 0.5, 2 and 5, as indicated by the labels on *x*-axis. Histograms are the calculated true emissivity of the designs generated by DT (color, fully opaque) and DTGEN (transparent black). Background light gray histogram shows the data used to train the models. Note: For clarity the scale of the training data distribution is different to the scale of the displayed y axis.

We test the spectral emissivity inverse design solutions for two types of cases. In case 1, we task the models with designing geometries for a specific purpose. For example in passive radiative cooling applications it is desirable to maximize the ratio of the emissivity inside a frequency band where the atmosphere is transparent over the emissivity outside this frequency band[3,4]. **Figure 5a** shows the leaf output from DTGEN that maximizes this metric, and the resulting spectra as calculated by DNS of the inverse designs, which all match the target spectra. In case 2, we randomly pick a sample from the test set and use it as the target emissivity spectrum for both DTGEN and DT. The results for case 2 are shown for each material type in **Figure 5b-d**. In all cases ~30 different designs are selected from the model-generated design rules, and the spectra for all designs are plotted (colored lines) against the target (black line). Because metal particles produce more diverse and complex spectra than dielectrics, the performance disparity between DTGEN and DT is most noticeable for the case of Au in **Figure 5b**. The model-generated designs produce the target emissivity spectra for dielectrics almost perfectly. **Figure 5e-f** show the rapid learning achieved with our models using comparatively small dataset sizes relative to other conventional machine learning models such as deep neural nets. It can be seen in **Figure 5e** that less training data is required for higher symmetry shapes (spheres > cylinders > parallelepipeds and triangular prisms), and more training data is required for metals. Points of different color and shape correspond to different particle materials and geometries, respectively. **Figure 5f** shows that DTGEN always outperforms DT for a given amount of training data, and uses the same point-labeling scheme as **Figure 5e**.

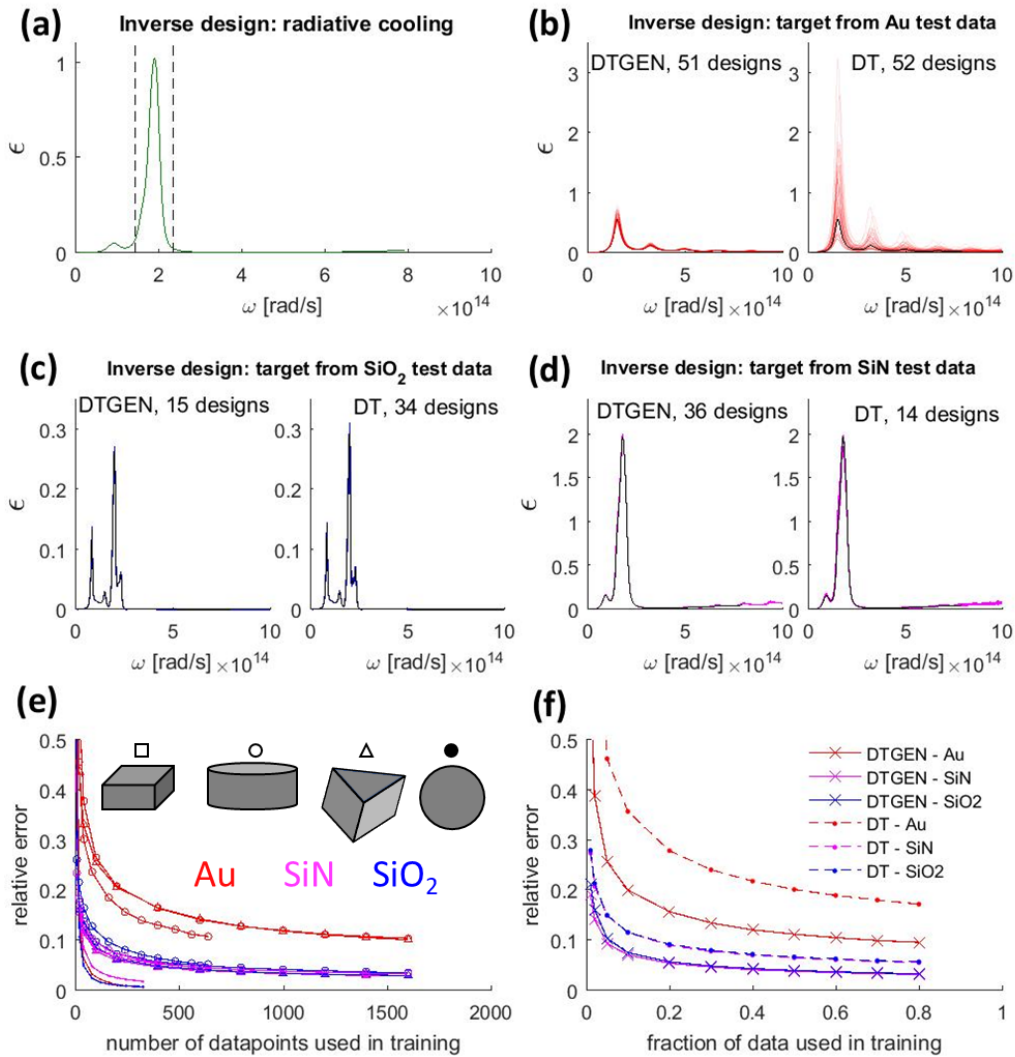

**Figure 5**. Inverse design solutions for spectral emissivity. For all spectra, black lines indicate the target optical properties and colored lines are the true optical properties of the model-generated designs. (a) Spectral emissivity of a DTGEN-generated design for maximize the ratio of the emissivity inside a frequency band where the atmosphere is transparent over the emissivity

outside this frequency band. This is situation is applicable to passive radiative cooling applications[3,4]. The ML model was tasked with maximizing the integrated emissivity over the atmospheric window (bounded by the black dashed lines) and minimizing it elsewhere. (b)-(d) compared spectra of DTGEN and DT generated designs for Au, SiO$_2$ and SiN, respectively. Targets randomly selected from test set. (e) Spectral emissivity learning curves for DTGEN models broken down by material and geometry (same colors and symbols as in **Figure 1** and **Figure 2**). Higher symmetry shapes require less training data; metals require more training data. (f) Spectral emissivity learning curves for DTGEN (or RF – slightly better performance) and DT broken down by material. Every point in the learning curves displayed in (e) and (f) represents the average of 100 independent training runs, with random test-train splits.

An RF model takes approximately 0.54 (8.4) ms/sample to train and 0.036 (0.227) ms/sample to predict the integrated (spectral) emissivity of a particle, using a single CPU. In contrast, numerically solving Maxwell's Equations to calculate the optical properties takes about 12 CPU-hours. As a surrogate model, a trained RF provides a $1.2\times10^9$ ($1.8\times10^8$) times speedup for the forward problem foe integrated (spectral) emissivity. For a training dataset size of 8,000 points RF still provides a 10,000 (640) times speedup for predicting one sample including the time it takes to train the RF. DT takes $4.7\times10^{-3}$ ($6.8\times10^{-2}$) ms/sample to train and 78 (120) ms/sample to solve the inverse design problem for integrated (spectral) emissivity targets. Assuming traditional optimization algorithms require 10 to 1000 iterations of solving the forward problem using DNS in order to solve the inverse problem, DT provides a $3.6\times10^6$ to $5\times10^8$ times speedup for the inverse problem, once trained. Including time to train on an 8,000 point dataset, DT provides a $6.5\times10^5$ to $3.7\times10^8$ speedup for inverse design. If improved accuracy and stability are desired for the inverse design, DTGEN takes approximately 11 (135) ms/sample to train, primarily due to generation of the synthetic dataset, and 0.2 (30) seconds to solve the inverse design problem for integrated (spectral) emissivity targets; still offering a $1.4\times10^4$ to $2.2\times10^8$ times speed-up once trained, or $1.4\times10^4$ to $1.5\times10^8$ times speedup including training time on an 8,000 point dataset. Runtimes are shown in **Fig. S12** and **Table S1**.

In conclusion, we have presented a method for creating interpretable machine learning models that can rapidly solve both the forward and inverse design problem for the spectral emissivity of particles of varying shape, size, aspect ratio, and material type. Future work will focus on experimental validation and extending our framework to be more generalizeable and incorporate design constraints. Machine learning is an exciting avenue for the discovery of novel optical metamaterial designs, and it offers unique learning and optimization capabilities.

## Methods:

### Numerical emissivity calculation

Hemispherical spectrally integrated emissivity and spectral emissivity of a finite particle with arbitrary shape is calculated by solving Maxwell's equations numerically. We use the fluctuating-surface-current formulation with the boundary-element method for its efficiency in directly calculating the integrated thermal radiation over all radiation directions in a single step[21]. In this method, the radiation emitted by the particle is calculated as a result of surface current fluctuations, which represent the random thermal motion of charges within the material. This method is implemented by a free and open source software, SCUFF-EM[22]. In the boundary-element method we create a surface mesh for the interfaces between any two distinct media, which in our case is the interface between the particle material and the surrounding vacuum. Fluctuating surface current is assumed at each mesh point, which is the source of thermal electromagnetic radiation. The radiated power at any point in space is calculated as the magnitude of the Poynting vector, which is then integrated over the entire surface of the finite particle to calculate the total radiated power. Thermal emissivity of the particle is defined as the ratio between the calculated total radiated power and the radiated power from a hypothetical blackbody at the same temperature multiplied by the same projected geometrical surface area as the particle of interest. We have created the surface meshes using gmsh, an open source software. We created the meshes to have facets that are small compared to the shortest wavelength (1.8 μm). We display the distribution of the mesh edge sizes in **Fig. S9**, which shows a mean value of 0.1632 μm for the square root of the surface area of the mesh surface facets. This is smaller than 10% of the shortest wavelength involved in our simulations. We performed a mesh convergence test to ensure the meshes were fine enough to achieve accurate results. Dielectric constants for Au[23], SiN[24] and SiO2[25] were adopted from literature.

## Analytical emissivity calculation

Geometries like infinitely wide thin films, infinitely long cylinders, and spheres, can be described using a single dimension in cartesian, cylindrical and spherical coordinates, respectively, due to their high symmetry. The spectral emissivities for these 1D geometries can be calculated analytically without the need for numerical solutions. For infinitely wide thin film, we used the transfer matrix method[26] to calculate the reflectance (R) and transmittance (T) of a single thin film, averaged between the two light polarizations (i.e., transverse electric and magnetic polarizations)[27]. Thermal emissivity is assumed to be equal to the absorbance of the film (A = 1 - R - T), based on Kirchoff's law.

$$\epsilon(\omega) = \frac{\int_{\theta=0}^{\pi} \epsilon(\theta, \omega) \sin\theta \cos\theta \, d\theta}{\int_{\theta=0}^{\pi} \sin\theta \cos\theta \, d\theta}$$

The emissivities of a sphere and a cylinder were calculated analytically using Mie theory[28,29]. Sphere hemispherical emissivity is calculated as the ratio between the absorption cross section and the sphere projected cross section[30]:

$$\epsilon_{sphere}(\omega) = \frac{\sigma_{abs,sph}(\omega)}{\pi r^2}$$

Sphere emissivity is independent of angle and light polarization due to the symmetry of the sphere. Cylinder directional emissivity is also calculated as the ratio between the absorption cross section and the cylinder projected cross section at a given angle, averaged between the two distinct light polarizations (i.e., electric field parallel or perpendicular to the plane formed by the cylinder axis and the radiation direction)[20]:

$$\epsilon_{cyl}(\omega, \theta) = \frac{\sigma_{abs,cyl}(\omega, \theta)}{2r \sin \theta}$$

Where $\sigma_{abs}$ is the absorption cross section of the infinite cylinder per unit length, and $\theta$ is the angle between the radiation direction and the cylinder axis. Hemispherical cylinder emissivity can be calculated from the directional emissivity from:

$$\epsilon_{cyl}(\omega) = \frac{\int_{\theta=0}^{\frac{\pi}{2}} \epsilon_{cyl}(\omega, \theta) r \sin^2 \theta \, d\theta}{\int_{\theta=0}^{\frac{\pi}{2}} r \sin^2 \theta \, d\theta}$$

Absorption cross sections for the sphere and cylinder were calculated as the difference between extinction and scattering cross sections.

### Dataset Description

We generated the majority of the dataset by random uniform sampling over the geometric parameters describing the particle for each material and geometry class: area-to-volume ratio (A/V), the shortest dimension (ShortDim), middle dimension (MiddleDim), and the longest dimension (LongDim). The range for A/V spanned from $10^6$ m$^{-1}$ to $10^8$ m$^{-1}$. We generated 15,900 datapoints including three materials: gold (Au), silicon dioxide (SiO$_2$), and silicon nitride (SiN); four geometry classes: sphere, cylinder, parallelepiped, and triangular prism; with 500, 800, 2000, and 2000 data points, respectively, for each material. We would like to emphasize that the 7,950 data points in our training set are considered sparse for this diverse combination of materials and geometries and the size of the resulting parameter space: 7,950/12=663 points per material + geometry class combination, and $663^{1/3} \cong 9$ points per linear dimension span, each of which ranges from 0.1 μm to 50 μm. We have shown the full distributions of geometry parameters in the supplementary section, **Figures S2-5.**

### Machine learning

Random Forests and Decision Trees were implemented in Scikit-learn[31]. For inference, we used random forest with 200 decision tree estimators. All experiments used a 50/50 train/test split ratio. We performed 100 random train-test splits with these ratios to provide estimates of the models' average performance.

We trained and tested our DT and RF models using 15,900 samples (7950 in training set, and 7950 in test set). For inverse design, we generated a synthetic (RF-labeled) dataset of 250×7950 = 1,987,500 data points, and used that combined with the original dataset to train DTGEN. The numerical emissivity calculations did not all sample at the same frequencies. Therefore we use spline interpolation to generate uniformly spaced 400-points-long emissivity spectra arrays. These 400 interpolation points were chosen in a logarithmic spacing from $10^{13}$ rad/s to $0.8×10^{14}$ rad/s.

All continuous input and output parameters were converted to log scale for training by taking the log of the inputs and then exponentiating the outputs. Log scale features result in more accurate and stable training and performance results for systems such as ours where parameters span several orders of magnitude. However, errors and loss functions were all calculated on a linear scale. While the models were trained by minimizing mean-squared-error (MSE) of log scale features, we report the relative error (of the linear scale emissivity) as the metric by which we evaluate model performance. Relative error is more appropriate than squared error for human-interpreted (linear scale) results because our emissivity data spans many orders of magnitude and MSE would disproportionately penalize errors of larger values for such linear scale features. We define relative error for integrated emissivity as $|\epsilon_{ML} - \epsilon_{DNS}|/\epsilon_{DNS}$, and for spectral emissivity as $E_{rel} = \frac{\int_\omega |\epsilon_{ML} - \epsilon_{DNS}| d\omega}{\int_\omega \epsilon_{DNS} d\omega}$.

**Acknowledgements**

This work was supported by the Laboratory Directed Research and Development Program (LDRD) at Lawrence Berkeley National Laboratory under contract # DE-AC02-05CH11231.

**Author Contributions**

M.E. conducted the numerical simulation, data generation, database construction, mass data visualization, performed literature search and applied ML methods for data analysis. C.Y. performed literature search, implemented/developed core code for RF, DT, DTGEN inverse design and inference, tested and validated inverse design methodology. The idea of using ML for inverse design was conceptualized by M.E., S.L. and R.P. and further developed by A.A. M.E, and C.Y. M.E., C.Y., S.L. and R.P, wrote the paper. S.L. and R.P supervised the research.

# Supplementary Figures ------------------------------------------------

**Analysis for emissivity dataset**

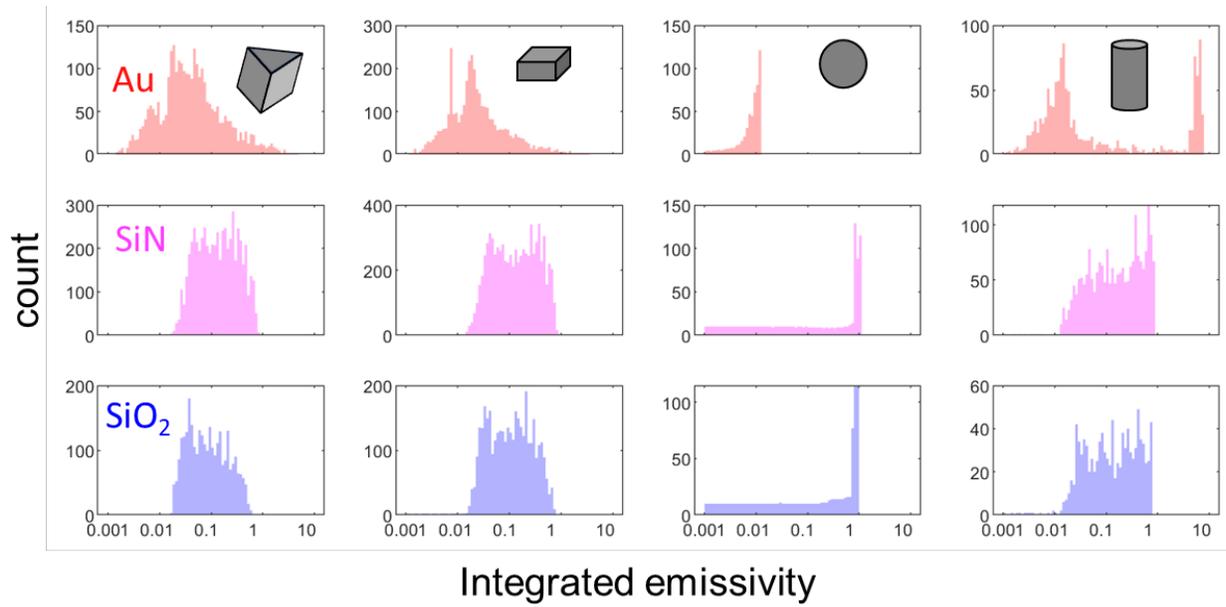

**Figure S1.** Integrated emissivity distribution broken by material (rows) and geometry (columns)

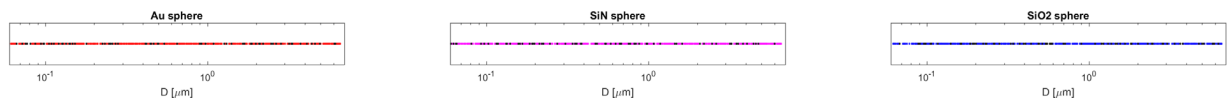

**Figure S2.** Input features for sphere. Test dataset are colored in black.

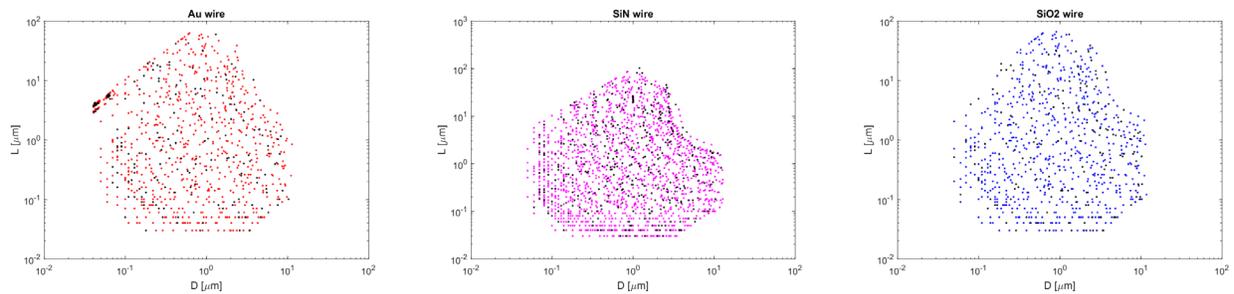

**Figure S3.** Input features for wires. Test dataset are colored in black.

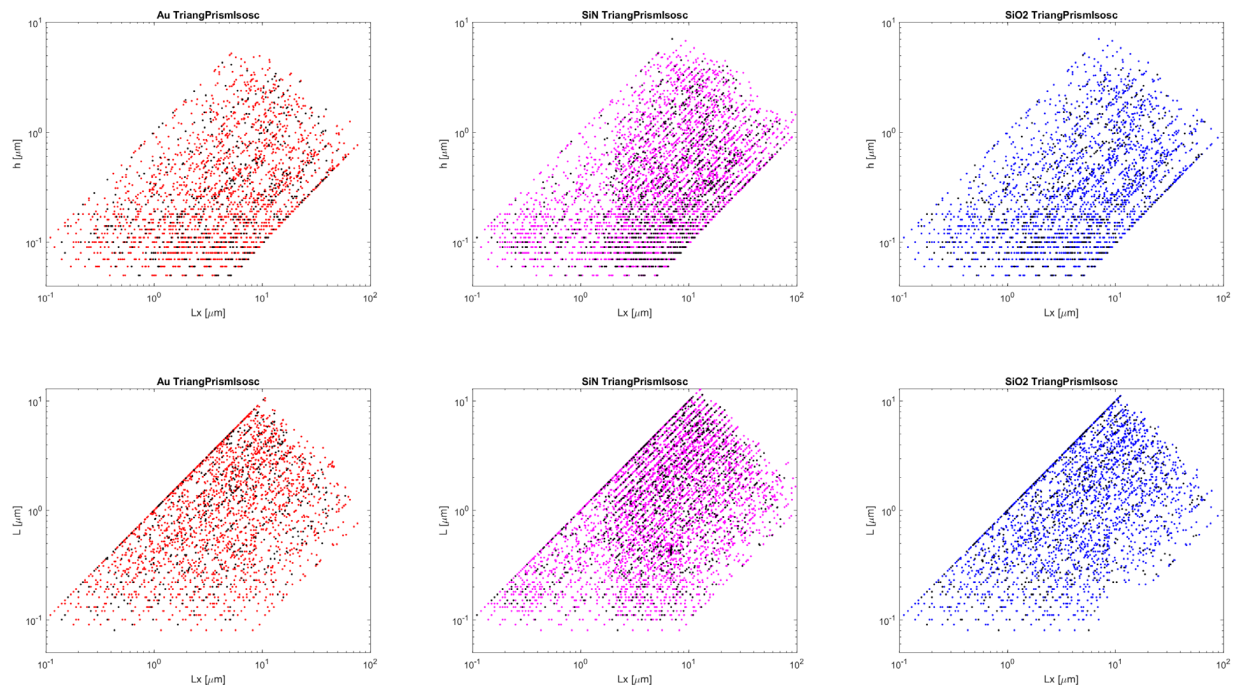

**Figure S4.** Input features for triangular prisms. Test dataset are colored in black.

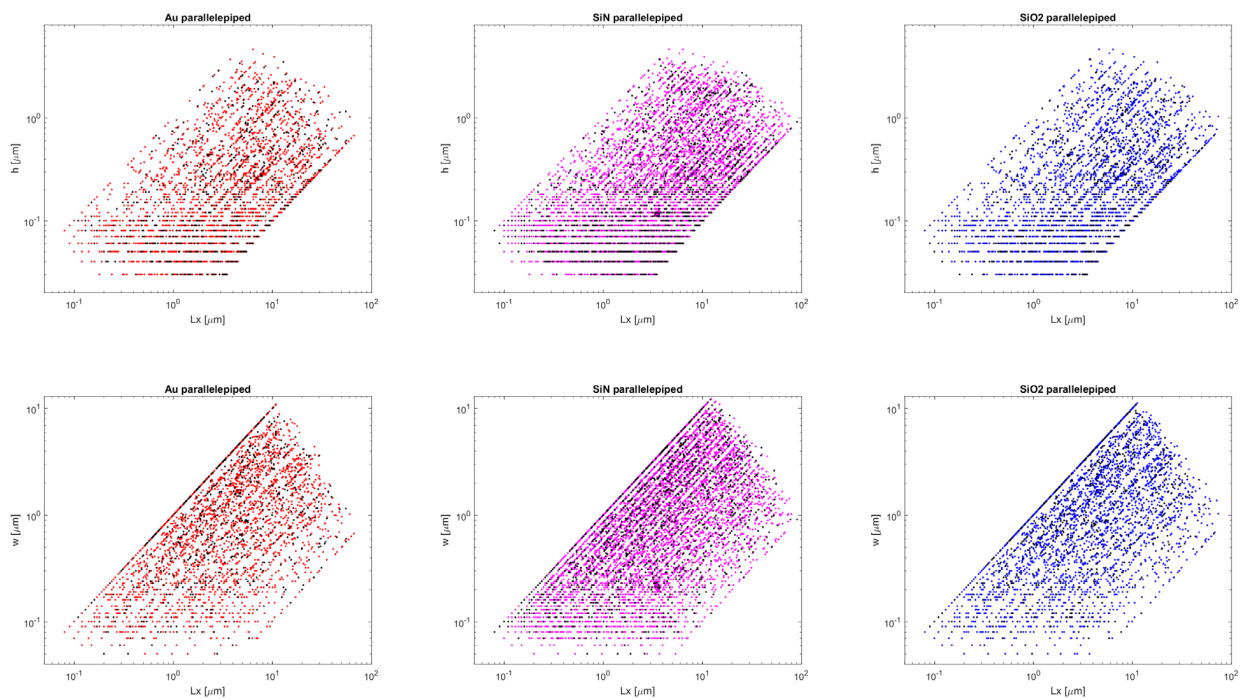

**Figure S5.** Input features for parallelepiped. Test dataset are colored in black.

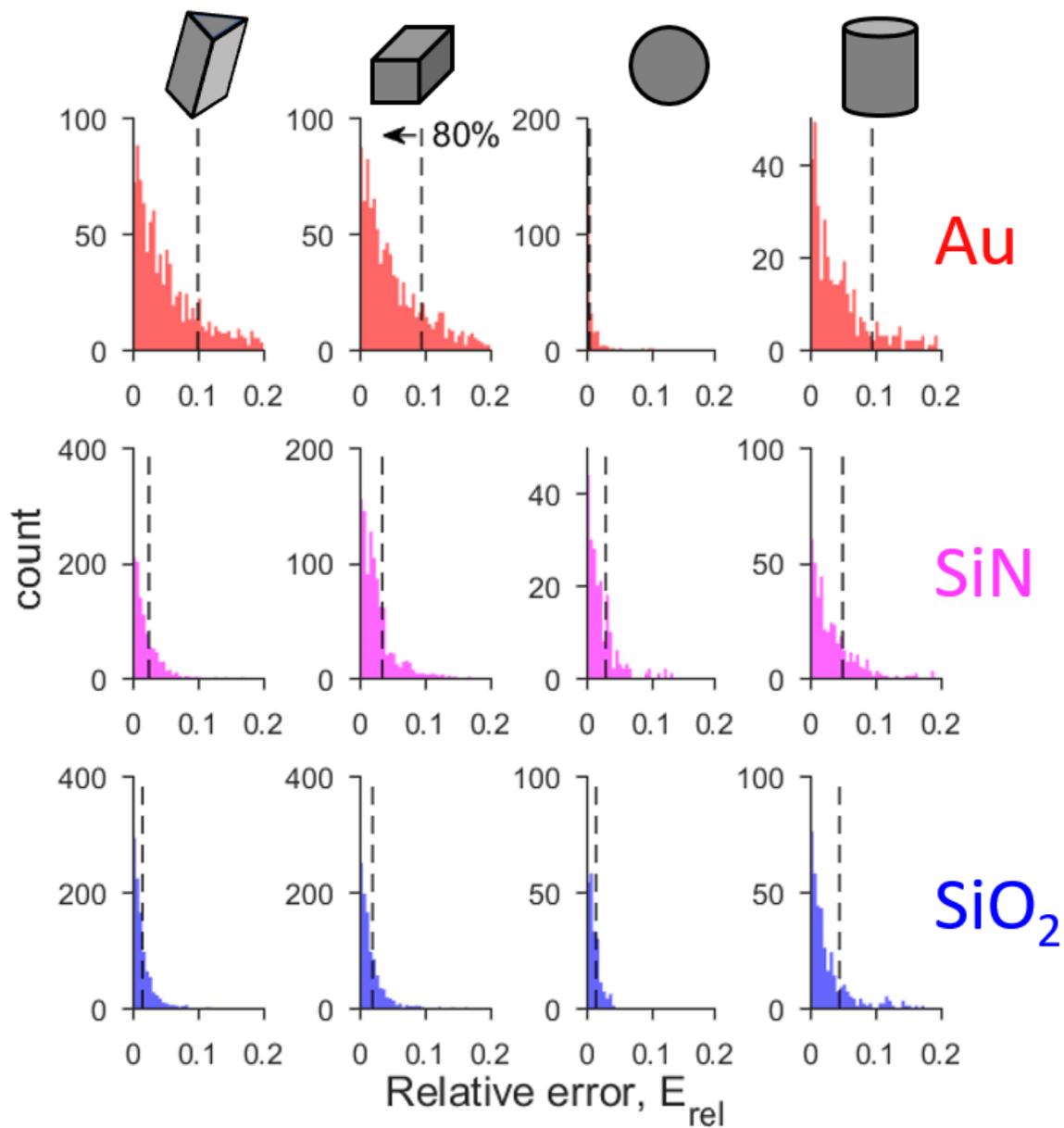

**Figure S6.** Relative error for machine learning model predictions for the integrated emissivity. The vertical broken line represents the edge of the 80% of the data.

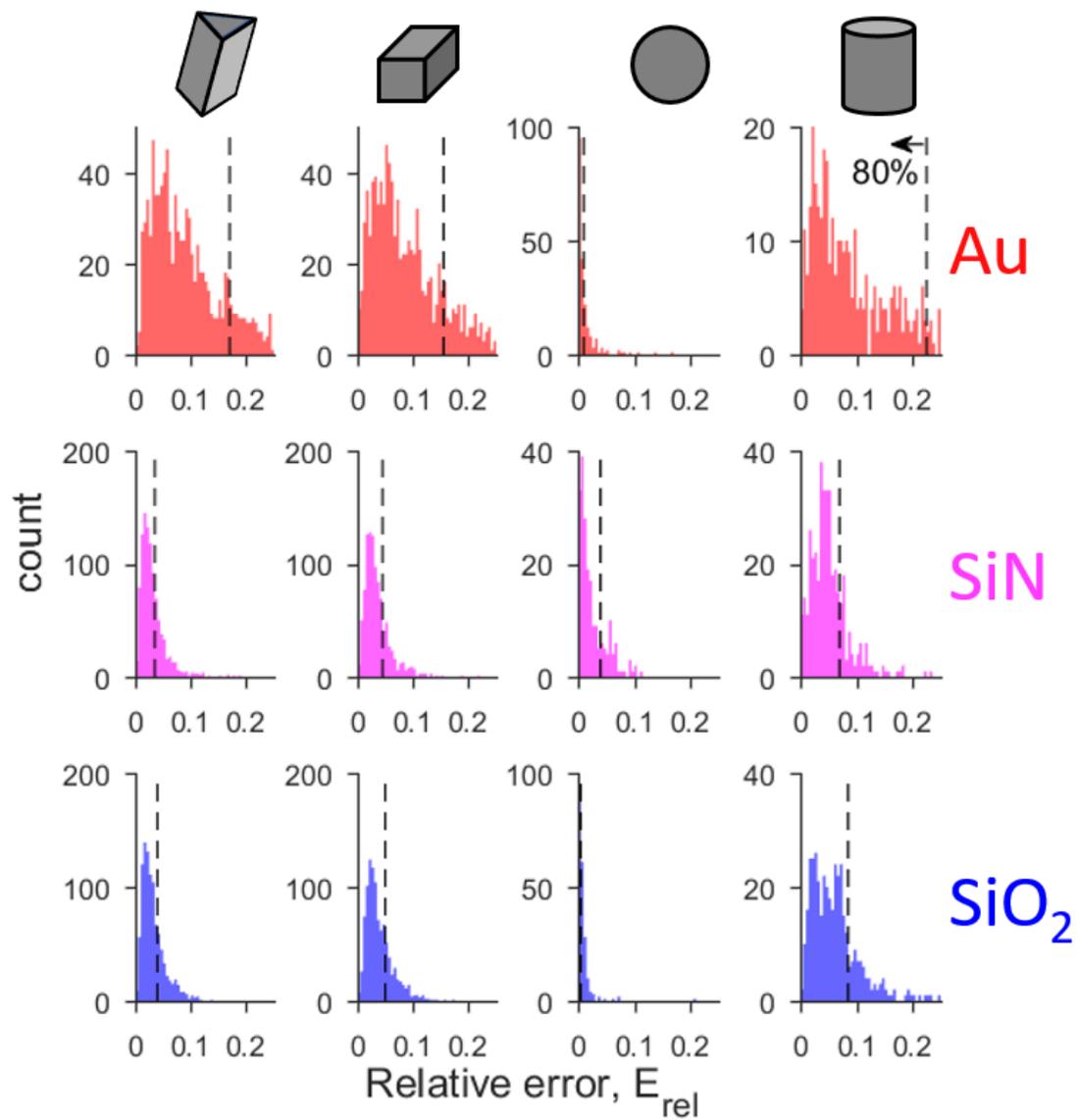

**Figure S7.** Relative error for machine learning model predictions for the spectral emissivity. The vertical broken line represents the edge of the 80% of the data.

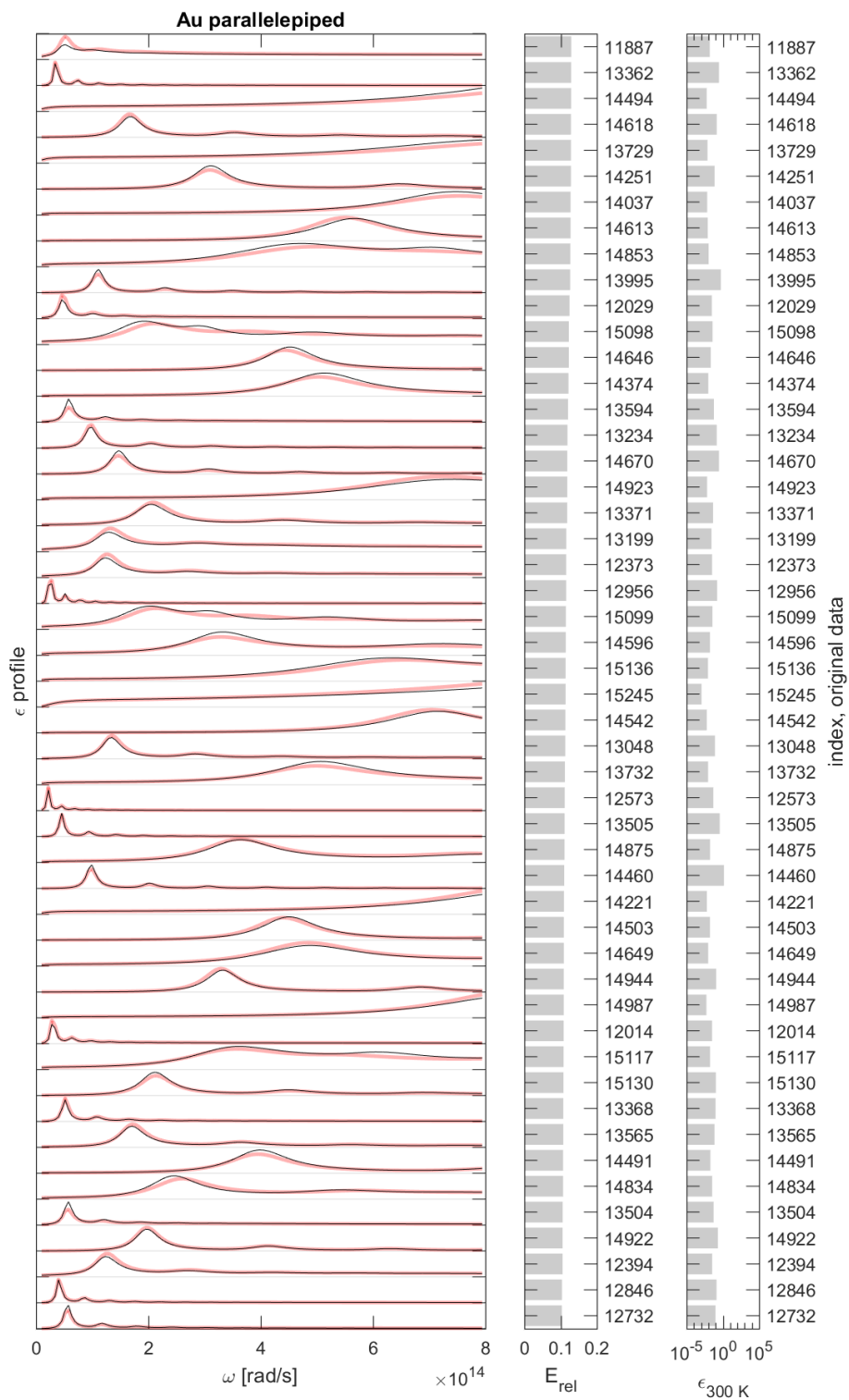

**Figure S8.** Machine learning prediction compared to numerical simulation for spectral emissivity for gold parallelepiped.

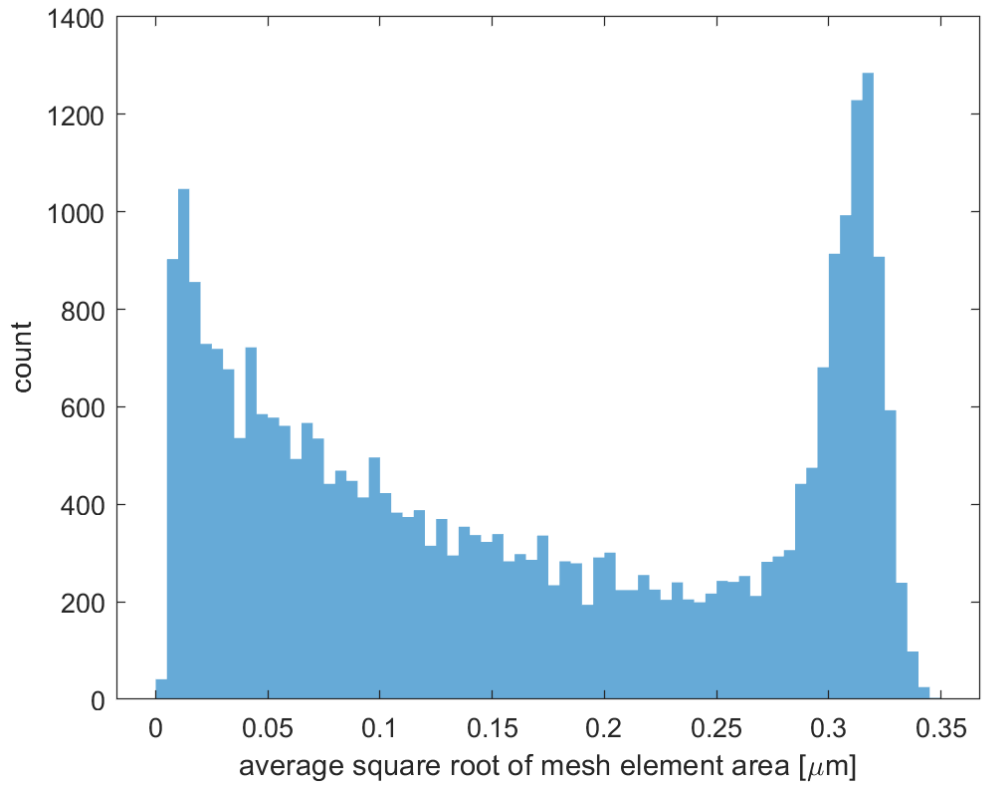

**Figure S9.** Distribution of the square root of the average surface area of the surface mesh elements for all the geometries. The mean value is 0.1632 µm.

# Design rules

| DT | DTGEN |
|---|---|
| **Design rules for achieving emissivity of 0.1, for the three materials** | |
| d_20191030_100754_DT_Material_Au_emiss_0.1.csv | d_20191030_100754_DTGEN_Material_Au_emiss_0.1.csv |
| d_20191030_100754_DT_Material_SiN_emiss_0.1.csv | d_20191030_100754_DTGEN_Material_SiN_emiss_0.1.csv |
| d_20191030_100754_DT_Material_SiO2_emiss_0.1.csv | d_20191030_100754_DTGEN_Material_SiO2_emiss_0.1.csv |
| **Design rules for achieving emissivity of 0.5, for the three materials** | **Design rules for achieving emissivity of 0.5, for the three materials** |

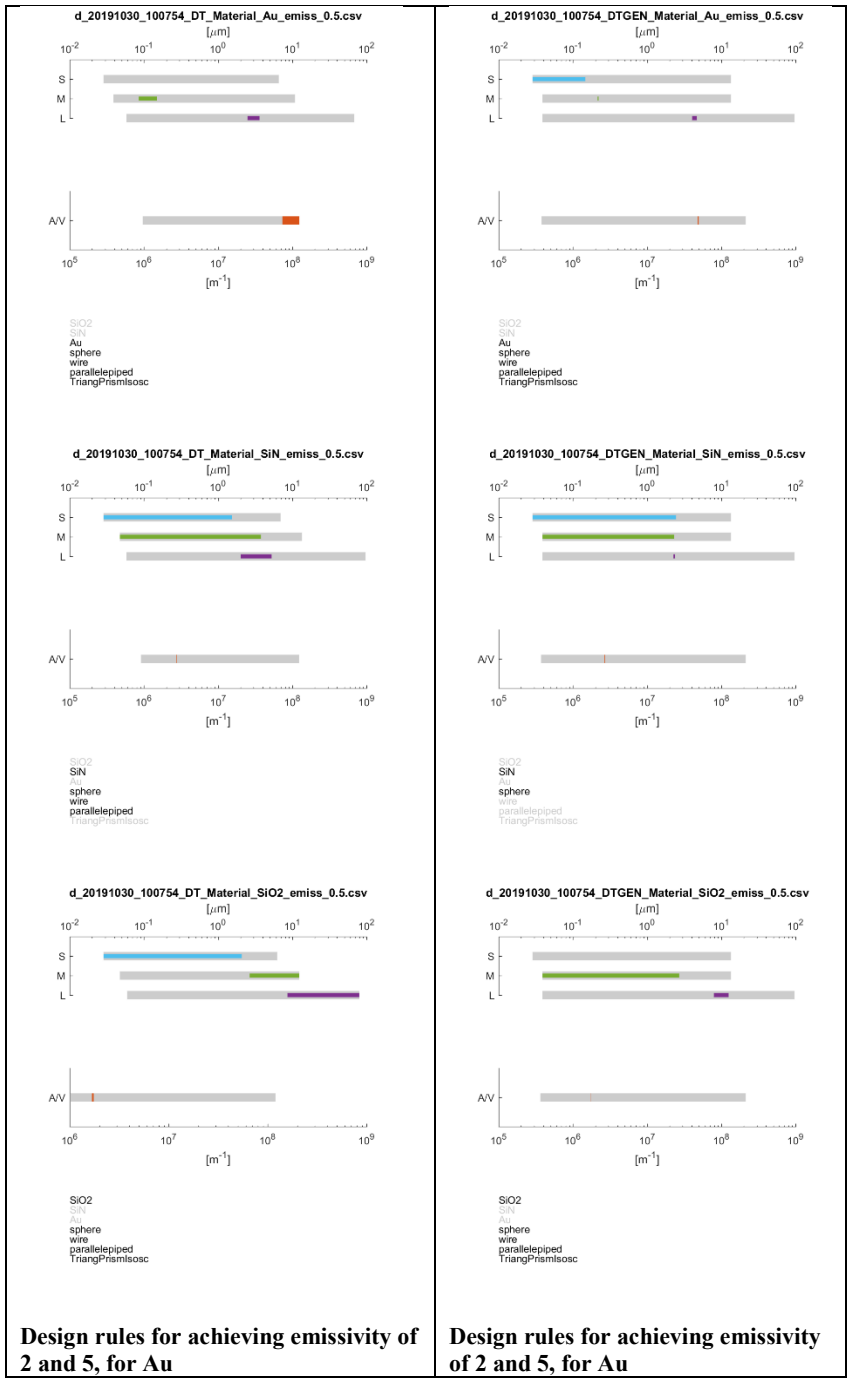

**Design rules for achieving emissivity of 2 and 5, for Au**

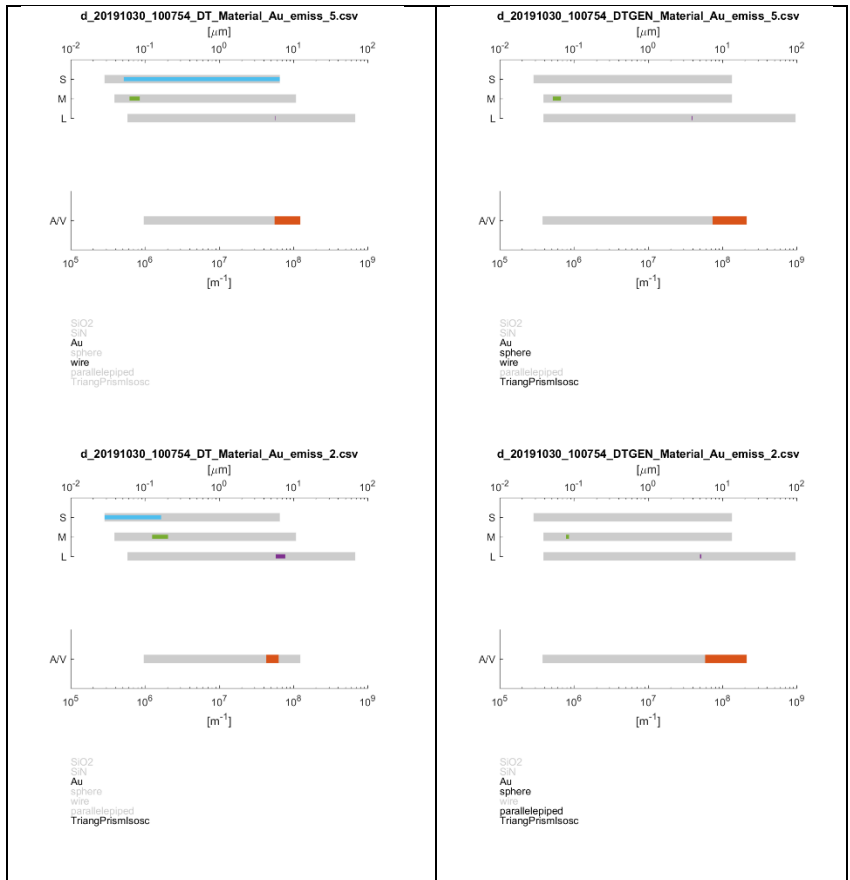

**Figure S10.** Design rules for integrated emissivity estimated by DT (left) and DTGEN (right). The grey shade represents the span of the parameter in the training data. For most cases, DTGEN shows stricter design rules than DT, which explains the superiority of DTGEN.

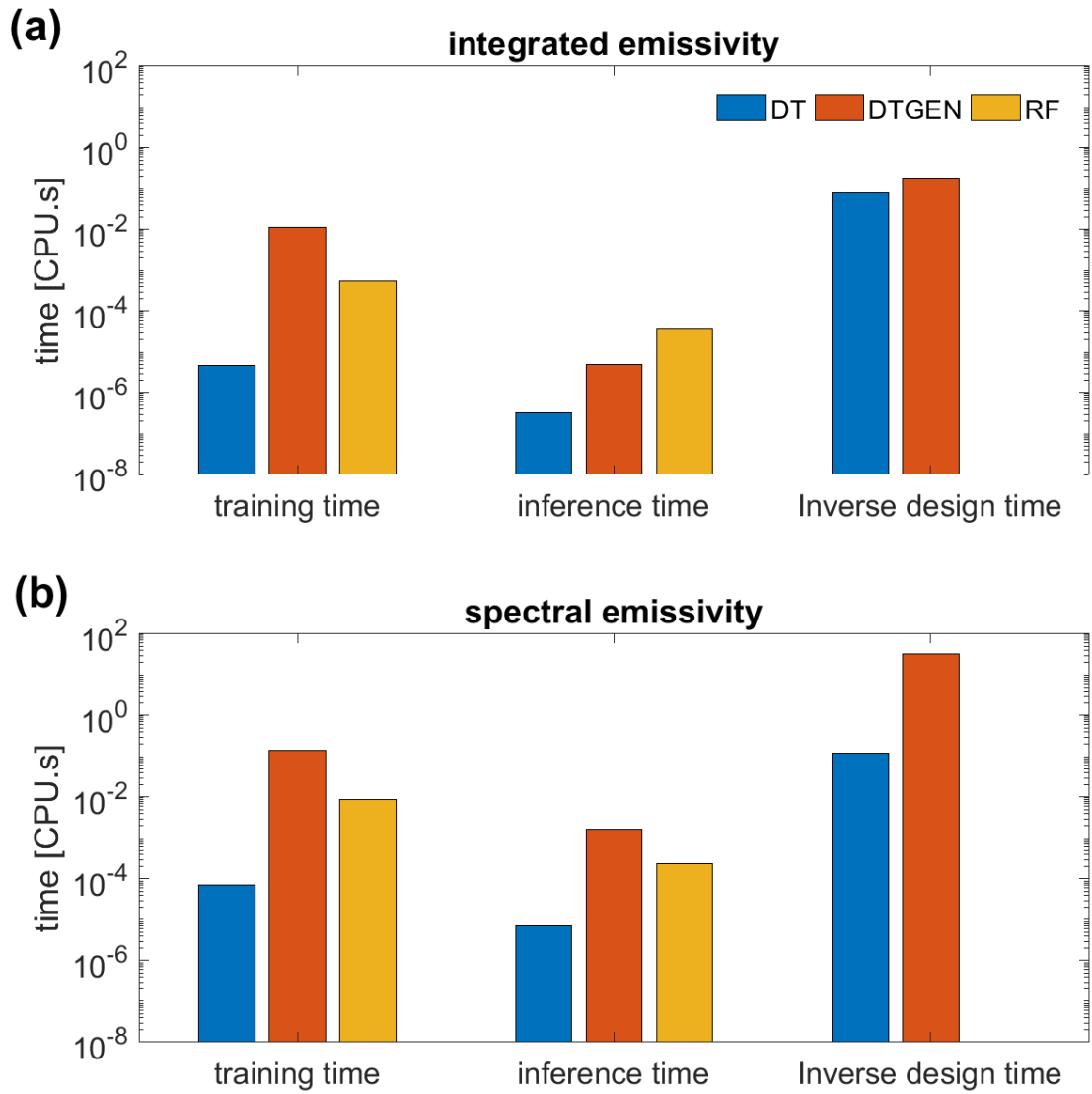

**Figure S11.** Training, inference and inverse design times for DT, DTGEN and RF models, per datapoint, using a single CPU. Average inference time for direct numerical simulation is 58 minutes using 12 cores.

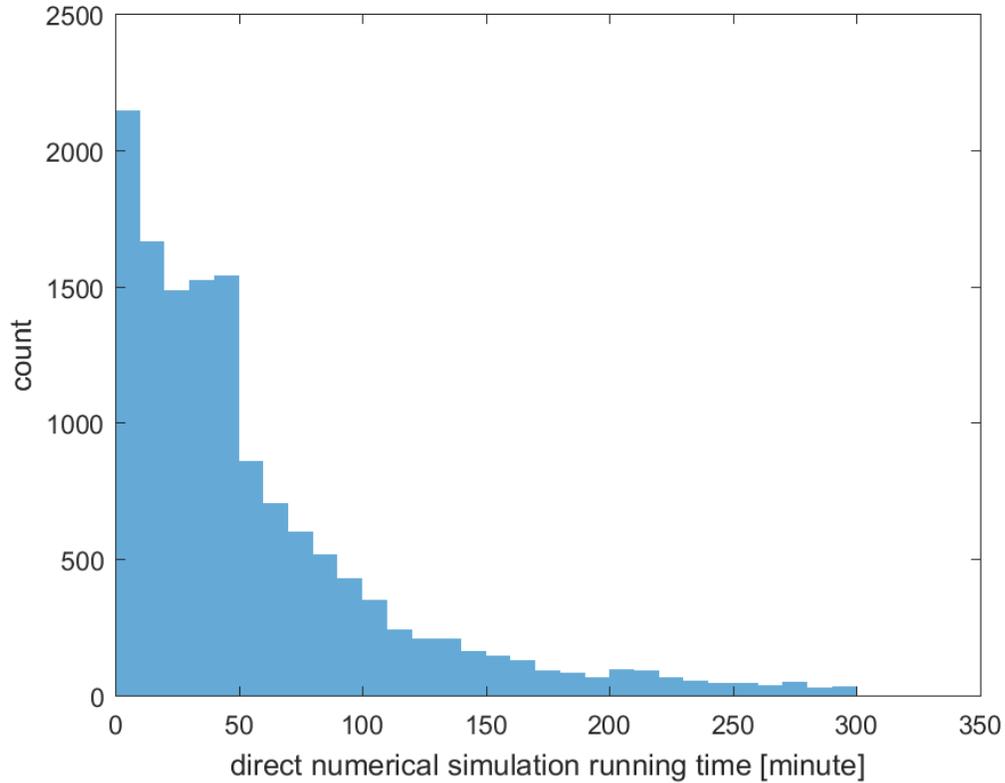

**Figure S12.** Direct numerical simulation wall clock time for training and test datasets using 12 cores. The average running time is 58 minutes.

**Table S1. Training, prediction and inverse design time for DT, RF and DTGEN models**

**Integrated emissivity**

|  | DT | RF | DTGEN |
|---|---|---|---|
| Training time (s) | 4.673e-06 | 5.363e-04 | 0.01103077 |
| Inference time (s) | 0.0032e-4 | 0.3553e-4 | 0.0486e-4 |
| Inverse design (s) | 0.078 | -- | 0.178289 |

**Spectral emissivity**

|  | DT | RF | DTGEN |
|---|---|---|---|
| Training time (s) | 6.8078e-05 | 0.0084 | 0.1350 |
| Inference time (s) | 6.865e-06 | 2.265e-04 | 0.00162 |
| Inverse design (s) | 0.1177 | -- | 31.31 |